\documentclass[fleqn,10pt]{wlscirep}

\usepackage{amsmath,amssymb}

\usepackage[T1]{fontenc}

\usepackage{textcomp,marvosym}

\usepackage{nameref,hyperref}

\usepackage{array}
\usepackage{sistyle}

\usepackage{graphicx}
\usepackage{epstopdf}

\newcommand{\be}{\begin{equation}}
\newcommand{\ee}{\end{equation}}
\newcommand{\beqn}{\begin{eqnarray}}
\newcommand{\eeqn}{\end{eqnarray}}

\title{Critical synchronization dynamics of the Kuramoto model on
  connectome and small world graphs}
\author[1]{G\'eza \'Odor}
\author[2,*]{Jeffrey Kelling}
\affil[1]{Institute of Technical Physics and Materials Science,
Centre for Energy Research,
P.O.Box 49, H-1525 Budapest, Hungary}
\affil[2]{Department of Information Services and Computing,
Helmholtz-Zentrum Dresden - Rossendorf,
P.O.Box 51 01 19, 01314 Dresden, German}

\affil[*]{j.kelling@hzdr.de}

\begin{abstract}
The hypothesis, that cortical dynamics operates near criticality also suggests,
that it exhibits universal critical exponents which marks the Kuramoto equation,
a fundamental model for synchronization, as a prime candidate for an underlying
universal model. Here, we determined the synchronization behavior of this model
by solving it numerically on a large, weighted human connectome network, containing
$\num{804092}$ nodes, in an assumed homeostatic state.
Since this graph has a topological dimension $d < 4$, a real synchronization
phase transition is not possible in the thermodynamic limit, still we could
locate a transition between partially synchronized and desynchronized states. 
At this crossover point we observe power-law--tailed synchronization durations,
with $\tau_t \simeq 1.2(1)$, away from experimental values for the brain.
For comparison, on a large two-dimensional lattice, having additional random, 
long-range links, we obtain a mean-field value: $\tau_t \simeq 1.6(1)$.
However, below the transition of the connectome we found global coupling
control-parameter dependent exponents $1 < \tau_t \le 2$, overlapping with 
the range of human brain experiments.
We also studied the effects of random flipping of a small portion of link
weights, mimicking a network with inhibitory interactions, and found similar
results. The control-parameter dependent exponent suggests extended dynamical 
criticality below the transition point. 
\end{abstract}

\begin{document}

\maketitle

\section*{Introduction}

Understanding the human brain, or in general neural systems is a great challenge of science,
in particular the application of models and methods of statistical physics has been
developing recently \cite{MArep}. 
There are several types of whole brain models, ranging from continuous,
integrate-and-fire models~\cite{Abbot89,DKJR} to discrete, activity spreading models~\cite{KH,Hai}.
All of them are effective ones, trying to describe different features of neural functions
measurable by neuroscience experiments. While different versions of integrating fire 
models are more detailed and use larger parameter space, simple activity spreading 
models try to capture basic features, like the emergence of power-laws (PL) of 
quantities via critical behavior \cite{Chi10}. Following experiments, these are
the neuron activity avalanche size and duration distribution tails, before finite
size cutoff. Criticality in these systems can be defined by the diverging correlation 
volume, as we tune a control parameter to a threshold value.

Criticality is an attractive hypotheses, because information processing and
dynamic range is optimal~\cite{Larr,KC}. 
Neural activity avalanche measurements have found power-laws, 
which arise naturally close to a critical point of a phase transition 
\cite{BP03,Fried,Shew,Yag,brainexp}.
The question of how a neural system would be tuned to this point has been debated. 
It was proposed to be by self-regulatory mechanisms \cite{stas-bak} leading to
self-organized criticality \cite{pruessner}, or as the consequence of extended 
dynamical critical regions in spreading models~\cite{MM,HMNcikk} in Griffiths
Phases (GP) \cite{Griffiths}.
The measured scaling exponents have been found to be close to the mean-field
transition values of discrete models \cite{odorbook}.

It is also known, that individual neurons emit periodic signals \cite{PSM16}, 
thus criticality may emerge by the collective behavior of oscillators
at the phase synchronization transition point. However, not much is known
about the dynamics of the synchronization or desynchronization process in
these models \cite{Pik,Acebron,CC18}.
Phase synchrony is essential for large-scale integration of information 
\cite{Varelaet,Buzs}, the role of the asynchronous state has remained 
more elusive \cite{Ren}. Very recently theoretical analysis of the homogeneous
Ginzburg-Landau type equations arrived at the conclusion that empirically 
reported scale-invariant avalanches can possibly arise if the cortex is
operated at the edge of a synchronization phase transition, where 
neuronal avalanches and incipient oscillations coexist \cite{MunPNAS}.

One of the most fundamental models, showing phase synchronization is the 
Kuramoto model of interacting oscillators \cite{kura}. This is defined on
full graphs, corresponding to the mean-field (MF) behavior \cite{chate_prl}, 
but as neural systems are not fully connected, we are interested in the 
phase synchronization transition in extended systems, where oscillators are
located at graph points, possessing finite topological dimension $d$. 
This is defined by
\begin{equation} \label{topD}
\langle N_r\rangle \sim r^d \ ,
\end{equation}
where $N_r$ is the number of node pairs that are at a topological
(also called ``chemical'') distance $r$ from each other
(i.e.\ a signal must traverse at least $r$ edges to travel from one
node to the other).  

Phase transition in the Kuramoto model can happen only above the lower 
critical dimension $d_l=4$ \cite{HPClett}. Below $d_l=4$ partial 
synchronization may emerge with a smooth crossover for strong coupling of 
oscillators, but a true, singular phase transition in the $N\to\infty$ 
limit is not possible. On higher dimensional full or random graphs the Kuramoto
equation exhibits universal
scaling dynamics of the phase order parameter~\cite{cmk2016,Kurcikk}. According
to the theory of
universality classes~\cite{odorbook} this "simple" model can describe 
other, more complex models of the brain. Very recently it has been studied
analytically and computationally on a human connectome graph network of 
$998$ nodes and in hierarchical modular networks (HMN), in which moduli 
exist within moduli in a nested way at various scales \cite{Frus}.
As the consequence of quenched, purely topological heterogeneity 
an intermediate phase, located between the standard synchronous and 
asynchronous phases was found, showing "frustrated synchronization", 
metastability, and chimera-like states. This complex phase was
investigated further in the presence of noise \cite{Frus-noise}
and on simplicial complex model of manifolds with finite and 
tunable spectral dimension \cite{FrusB} as simple models for the brain.

Anatomical connections~\cite{SCKH04} and the synchronization networks
of cortical neurons~\cite{22} indicate a small-world topology~\cite{WS98}.
Here we will investigate the characteristic times, corresponding to
synchronization or desynchroziation near the transition on a large human
connectome graph {\it (KKI-18)} and compare it with results, obtained on 
2d lattices with additional random, long range connections. The
latter also exhibit small-world topology, because the 2d lattice has
large clustering and the random, long range connections generate
short path lengths among geometrically distant nodes.
Our graphs are much larger than those considered before, allowing us to
determine universal critical exponents that can be compared with experiments.
Furthermore, we have heterogeneity in the intrinsic frequencies as well as
in connection weights, which was found to be crucial in case of 
threshold model simulations~\cite{CCdyncikk}.
Previously, extended discrete threshold model simulations of activity
avalanches on {\it KKI-18} did not support a critical phase 
transition~\cite{CCdyncikk}.
It turned out the weight heterogenities were too strong to allow the
occurrence of criticality. This means that only the strongly connected
hubs played a role in the activation/deactivation processes and weak
nodes just followed them. As this appears to be unrealistic and
uneconomic in a brain of billions of neurons, an input sensitivity
equilibration was assumed via variable, node dependent thresholds.
This makes the system homeostatic and simulations proved the occurrence of
criticality, as well as robust Griffiths effects~\cite{CCdyncikk,CC-tdepcikk} in
spreading models.
Indeed, there is some evidence that neurons have a certain adaptation to their
input excitation levels~\cite{neuroadap} and can be modeled by variable
thresholds~\cite{thres}.
Very recently comparison of modeling and experiments arrived at
a similar conclusion: equalized network sensitivity improves the
predicting power of a model at criticality in agreement with
the FMRI correlations \cite{Rocha2008}.

Even more naturally, homeostasis can be achieved in real brains 
via inhibitory neurons \cite{Homeo-inh,65,66,67,68}, suppressing communications. 
This provides an alternative way for modifying the positive, undirected links of the
{\it KKI-18} graph to test the phase synchronization of the Kuramoto model
in the presence of random, negative couplings.

\section*{Models and methods}

We consider the Kuramoto model of interacting oscillators \cite{kura}, with phases $\theta_i(t)$ 
located at $N$ nodes of networks, which evolve according to the 
dynamical equation
\be
\dot{\theta_i}(t) = \omega_{i,0} + K \sum_{j} W_{ij} \sin[ \theta_j(t)- \theta_i(t) ]
\label{kureq}
\ee
Here, $\omega_{i,0}$ is the intrinsic frequency of the $i$-th oscillator, drawn from a 
Gaussian distribution with zero mean and unit variance and the summation 
is performed over other nodes, with connections described by the weighted 
adjacency matrix $W_{ij}$. The global coupling $K$ is the control parameter
of this model, by which we can tune the system between asynchronous and synchronous states.
We follow the properties of the phase transition through studying the Kuramoto 
order parameter defined by
\be
R(t)=\frac{1}{N}\left|\sum_{j=1}^Ne^{i\theta_j(t)}\right|,
\label{op}
\ee
which is non-zero, above a critical coupling strength, $K > K_c$ tends to 
zero for $K < K_c$ as $R \propto\sqrt{1/N}$ or exhibits a growth at $K_c$
as
\be
R(t,N) = N^{-1/2} t^{\eta} f_{\uparrow}(t / N^{\tilde z}) \ ,
\label{escal}
\ee
in case of an incoherent initial state, with the dynamical exponents ${\tilde z}$ 
and $\eta$.
In case of a coherent initial state it decays as:
\be
R(t,N) = t^{-\delta} f_{\downarrow}(t / N^{\tilde z}) \ ,
\label{dscal}
\ee
characterized by the dynamical exponent $\delta$. Here $f_{\uparrow}$ and
$f_{\downarrow}$ denote different scaling functions.

We have also investigated the de-synchronization duration distributions by starting
the system from fully synchronous or asynchronous states, near $K_c$ 
by measuring the time $t_x$ until $R(t_x)$ first fell below the threshold value:
$R_T=1 / \sqrt{N}$, related to the synchronization noise in the incoherent phase
(see Fig.~\ref{figR}).
For this measurement we averaged over $\simeq 10^4$ runs, using independent
random $\omega_{i,0}$ intrinsic frequencies and applied histogramming with
increasing bin sizes: $\Delta t_x \propto t_x^{1.12}$ to estimate the 
probability distribution $p(t_x)$.

The following graphs have been considered:
\begin{enumerate}
\item 2d lattices with additional, random long-range connections such that
$\langle k\rangle = 5$ (2dll). We used periodic boundary conditions, simulating 
high dimensional graphs with supposedly mean-field behavior.
\item Weighted, symmetric large human connectome graph: {\it KKI-18} \cite{CCcikk}
downloaded from the Openconnectome project \cite{OCP}. 
\item {\it KKI-18}, with 5\% of the links turned to inhibitory:  {\it KKI-18-I}
\end{enumerate}

We applied the fourth order Runge-Kutta method from Numerical Recipes~\cite{NumR}
and the boost library odeint~\cite{boostOdeInt} to solve Eq.~(\ref{kureq}) 
on various networks. Step sizes: $\Delta = 0.1, 0.01, 0.001$
have been tested, but finally  $\Delta =0.1$ precision found to be sufficient. 
Generally, the $\Delta < 0.1$ precision did not improve the stability of the
solutions, but caused smaller fluctuations due to the chaotic behavior of Eq.~(\ref{kureq})
which could be compensated by averages over many independent samples with
different $\omega_{i,0}$.
The criterion $\epsilon = 10^{-12}$ was used in the RK4 algorithm and
we parallelized the RK4 for NVIDIA graphic cards (GPU), by which we could
achieve a $\sim\times 40$ increase in the throughput with respect to a 
single 12-core CPU. It is important to note, that first we verified the 
parallel GPU code, by comparing results on smaller sizes with those, 
obtained by the serial CPU program.
Algorithmic and benchmark details will be discussed elsewhere \cite{GPU-kur}.

We measured the Kuramoto order parameter with a fixed $K$, by increasing the sampling time
steps exponentially
\begin{equation}
t_k = 1 + 1.08^{k} \ ,
\end{equation}
which is a common method in case of PL asymptotic time dependences. In practice
we estimate $t_x = (t_k + t_{k-1})/2$, where $t_k$ was the first measured
crossing time.
The initial conditions were generally $\theta_i(0) \in (0,2\pi ]$ phases, with uniform
distribution, describing fully disordered states. However, for comparison we 
also performed runs starting from the fully synchronized state: $\theta_i(0) = 0$. 
Probability distribution tails were fitted using the least squares fit method 
above thresholds, fixed by visual inspection of the results.
To see the corrections to scaling we determined the effective exponents
of $R$ as the discretized, logarithmic derivative of Eq.~(\ref{escal}) at
these discrete timesteps $t_k$, near the transition point
\begin{equation}  \label{Reff}
\eta_\mathrm{eff} = \frac {\ln \langle R(t_{k+3})\rangle - \ln \langle R(t_{k})\rangle} 
{\ln(t_{k+3}) - \ln(t_{k})} \ .
\end{equation}
Here the brackets denote sample averaging over different initial conditions.

\begin{figure}[!h]
\includegraphics[height=5.5cm]{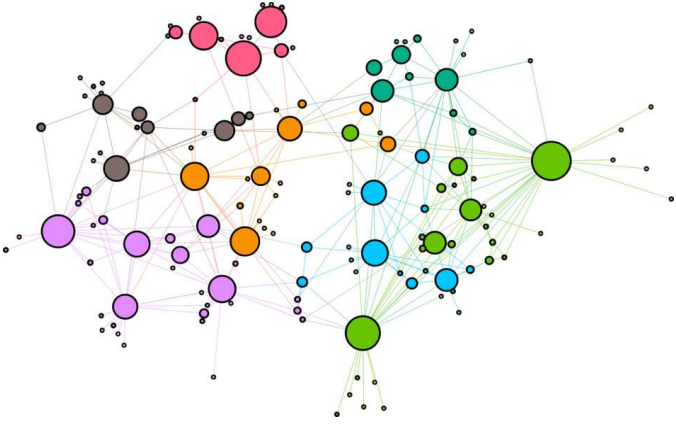}
\caption{Network of the modules of the {\it KKI-18} human connectome graph.
The size of circles is proportional with the number of nodes.}
\label{modules}
\end{figure}

The {\it KKI-18} graph has been downloaded from the Open Connectome project 
repository \cite{OCP}. This network was generated from Diffusion Tensor
Imaging (DTI)~\cite{DTI}, approximating the {\it structural connectivity}
of the white matter of a human brain.
It comprises $N=\num{804092}$ nodes, connected via \num{41523908} undirected edges, 
and several small sub-components, which were ignored here.
~\footnote{Note, that keeping the sub-components, did not change the results 
within numerical accuracy.}.
Solving (\ref{kureq}) on this graph allows running extensive dynamical
studies on present day CPU/GPU clusters, large enough to draw conclusions on 
the scaling behavior without very strong finite size effects.
These connectomes of the human brain possess \SI1{mm^3} resolution,
using a combination of diffusion weighted, functional and structural
magnetic resonance imaging scans. They are symmetric, weighted
networks, where the weights measure the number of fiber tracts between nodes.
The large graph "KKI-18" used here is generated by the MIGRAINE method,
described in \cite{MIG}. They exhibit hierarchical levels by construction 
from the Desikan cerebral regions with (at least) two quite different scales. 
The graph structure can be seen on the Fig.~\ref{modules}, where the modules 
were identified by the Louvain algorithm~\cite{louvainAlgo},
then the network of modules was generated and finally visualized 
using the Gephi tool~\cite{gephi}. This identified $144$ modules, 
with sizes varying between $8$ and $\num{35202}$ nodes. 

In~\cite{CCcikk} it was found that, contrary to the small world network
coefficients, these graphs exhibit topological dimension slightly above
$D=3$ and a certain amount of universality, supporting the selection of
\textit{KKI-18} as a representative of the large human connectomes available.
This dimensionality suggests weak long-range connections, in addition to
the $D=3$ dimensional embedding and warrants to see heterogeneity
effects in dynamical models defined on them.

To keep the local sustained activity requirement for the brain \cite{KH} 
and provide a homeostatic state, we modified  {\it KKI-18} by normalizing the
incoming weights of node $i$ in \cite{CCdyncikk}:
$W'_{i,j} = W_{i,j}/\sum_{j \in {\rm neighb. of} \ i} W_{i,j}$
at the beginning of the simulations.

In addition, it is well known that excitation is balanced globally by the 
inhibitory cells, which is assumed to be a $20\%$ fraction of all neurons.
However, the vast majority of inhibitory connections are local as these 
cells have small dendritic and axonal trees \cite{Anatomy}.
Their range coincides roughly with the $1mm^3$ of voxels in the connectivity 
data of {\it KKI-18}.
The tracts, obtained by DTI are in contrast mostly excitatory axonal fibers; 
middle and long range connections are made by axons of pyramidal cells, 
which are excitatory. But these axons still could target excitatory and
inhibitory cells in a voxel's area.
To model this, we flipped the signs of weights of $5\%$ randomly selected 
links $i$ as $W''_{i,j} = -W'_{i,j}$, creating thus a modified graph called: 
{\it KKI-18-I}. Such links are against local synchronization and can be
considered as an inhibition mechanism of possible information retrieval 
mechanism via resonance \cite{ResX}.

\section*{Results}

\subsection*{The 2dll graph}

\begin{figure}[!h]
\includegraphics[height=5.5cm]{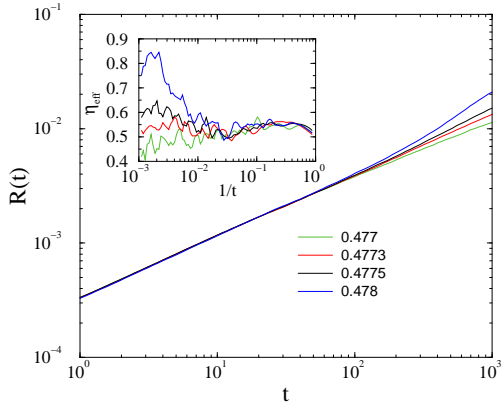}
\caption{Growth of the average $R$ on the 2dll model near the synchronization 
transition point for $K=0.477, 0.4773, 0.4775, 0.478$ (bottom to top curves).
Inset: the corresponding local slopes defined by (\ref{Reff}).}
\label{figll}
\end{figure}

First we studied the growth of $R(t)$ on the 2dll model of linear size $L=6000$
by starting from states of oscillators with fully random phases
and by averaging over $5000-10000$ $\omega_{i,0}$ realizations up to $t = 10^3$. 
As Fig~\ref{figll} shows power-law growth of synchronization emerge up to 
$t\simeq 100$ in the coupling region: $0.477 \le K \le 0.478$. 
Following that the $R(t)$ curves veer up or down, depending on being super 
or sub-critical. Note, that for $t > \sim 800$ the curves begin to 
break down, owing to the finite size effect, when the growing correlation 
volume: $\xi\propto t^{\tilde z}$ exceeds the system size $N=L^2$.
Looking at the effective exponents defined by (\ref{Reff}) one can estimate 
the critical point: $K_c = 0.4775(3)$, as we expect in the asymptotic 
$1/t\to\infty$ limit constant local slopes. Off-critical cases exhibit
up or down veering curvatures.
Here one can read-off the asymptotic value: $\eta = 0.55(10)$, on the local 
slope inset of Fig~\ref{figll}, which is different from the MF value $\eta = 0.75$,
expected for the Kuramoto model \cite{cmk2016} by scaling relations. 
The obvious discrepancy can be the consequence
of very strong corrections-to scaling or some other quench disorder effect
discussed further in \cite{Kurcikk}. Note, that in \cite{cmk2016},
in case of fully connected graphs, the $\eta = 0.75$ exponent could hardly be seen,
probably as the consequence of finite size and time limitations. While \cite{cmk2016}
achieved sizes $N \le \num{819200}$, here we provide results for much larger systems, 
containing $N=\num{36000000}$ nodes, but without full topological order.

\begin{figure}[!h]
\includegraphics[height=5.5cm]{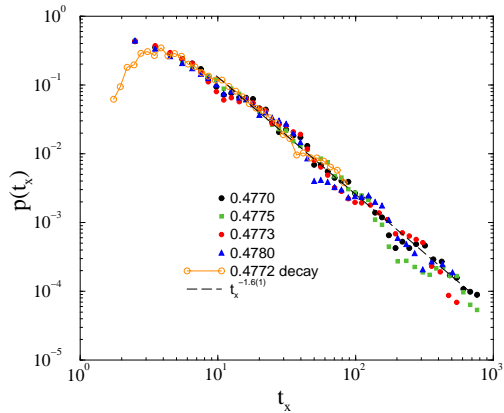}
\caption{Duration distribution of $t_x$ on the 2dll model for growth with $L=6000$:
$K=0.477$ (bullets), $0.4773$ (boxes), $0.4775$ (diamonds), $0.478$ (triangles)
and decay with $L=1000$ for $K=0.4772$ (solid line).
The dashed line shows a PL fit to the $K=0.477$ line for the tail region: $t_x > 10$.}
\label{durll}
\end{figure}

Instead of going into the details we just show that the relaxation time 
distributions, both for growth and for decay result in a robust 
fat tail behavior at the critical point, characterized by $\tau_t=1.6(1)$
(see Fig.~\ref{durll}).
The evolution of single realizations, in case of growth runs are shown 
on Fig~\ref{figR}. The dashed line denotes the threshold at which the
first passage time $t_x$ is measured. 
We have also measured $t_x$ in the case of fully coherent initial states
in systems up to $t_\mathrm{max}=10^4$.
Fig.~\ref{durll} shows the tails of $p(t_x)$ around the critical point 
for incoherent initial conditions for $L=6000$ and for coherent initial
state with with $L=1000$. In the latter case the decay occurs following a
long transient time, thus the time is divided by a factor $100$, but one
can observe the same type of PL tails at $K \simeq 0.477$.  

With this precision one cannot see a difference in the numerical scaling behavior
at the critical point, but we note that the $\tau_t=1.6(1)$ estimate is slightly 
above that one could obtain using the scaling relation
\be
\tau_t = 1 + \delta = 1.5 \ ,
\ee
connecting dynamical exponents, see for example \cite{odorbook}. 
Note, that the $\delta = 0.6(1)$ result is in agreement with those 
of the extensive decay results, presented in \cite{Kurcikk} and suggest that 
$\delta \simeq \eta$ on this scale, ruling out possible strong
artifacts of the threshold value selection.

\subsection*{The Connectome graph}

\subsubsection*{Normalized, positive weights}

\begin{figure}[!h]
\includegraphics[height=5.5cm]{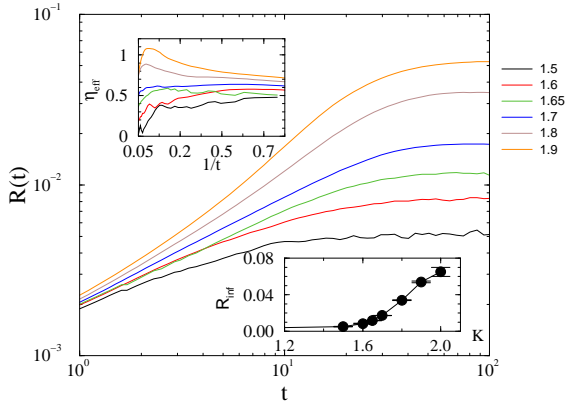}
\caption{Growth of the average $R$ on the {\it KKI-18} graph near the synchronization
transition point for $K=1.5$, $1.6$, $1,65$, $1.7$, $1.8$, $1.9$ (bottom to top curves).
Upper left inset: Effective exponents, defined by (\ref{Reff}) for the same data,
down inset: steady state $R(t\to\infty)$ as the function of the global coupling.}
\label{figccgrowth}
\end{figure}

In case of the {\it KKI-18} graph first we determined the crossover point via
the inflexion condition, which separates up (convex) and down (concave)
veering curves of the growth runs (see Fig~\ref{figccgrowth}). 
As we can see, the transition is much smoother than what we obtained
in the 2dll graph.
The lower inset of Fig~\ref{figccgrowth}) shows the steady state values
$R(t\to\infty)$ as the function of $K$, with a very low level of 
synchronization above the transition. This smooth crossover behavior
is not surprising, as the topological dimension of this graph is: 
$d = 3.05 < d_l = 4$ \cite{CCcikk}.
This behavior is in agreement with PET and fMRI studies, which suggest that 
the magnitude of activity change from rest to task is rather small.
Looking at the shapes of the $R(t)$ curves and the saturation level of the
corresponding local slopes we can estimate this crossover at: $K_c = 1.65(5)$, 
with an effective scaling exponent $\eta_\mathrm{eff} \simeq 0.6(1)$.
This exponent value is smaller than the $\eta = 0.75$ MF value
for the Kuramoto model, but close to our 2dll graph results.

\begin{figure}[!h]
\includegraphics[height=5.5cm]{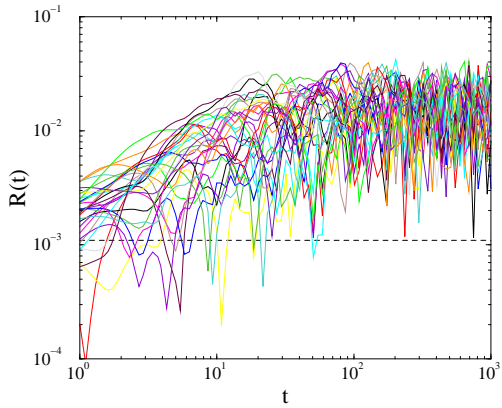}
\caption{Evolution of $R(t)$ for single realizations on the {\it KKI-18} graph
at $K=1.7$. 
The dashed line shows the threshold value $R = 1/\sqrt(N) = 0.001094$, 
where we measure the characteristic times: $t_x$ of first cross.}
\label{figR}
\end{figure}

\begin{figure}[!h]
\includegraphics[height=5.5cm]{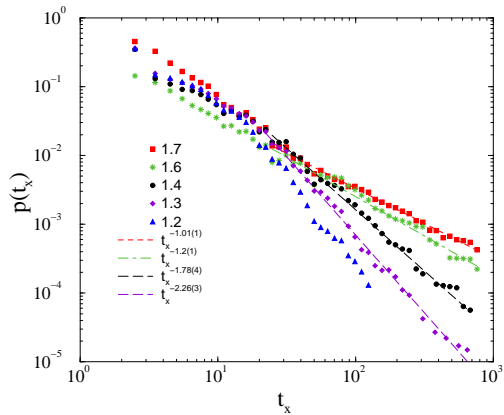}
\caption{Duration distribution of $t_x$ on the {\it KKI-18} model for growth
$K=1.7$ (boxes), $1.6$ (stars), $1.4$ (bullets),  $1.3$ (+), $1.2$ (up triangles).
The dashed lines shows PL fits for the tail region $t_x > 20$.}
\label{durcc}
\end{figure}

Having determined the transition point we run the numerical solver 
at control parameter values below $K_c$, by starting with thousands 
of random initial states and measuring the first crossing times 
$t_x$, when $R$ fell below: $1/\sqrt(N) = 0.001094$ (see Fig.~\ref{figR}).
Following a histogramming procedure, with PL growing bin sizes in $t_x$, 
we obtained the distributions $p(t_x)$, which exhibit PL tails, 
characterized by the exponents $1 < \tau_t < 2$ (see Fig.~\ref{durcc}). 
Here the almost $\sim 1/t$ decay at $K_c \ge 1.7$ marks synchronized phase,
with singular behavior; at $K_c$ one obtains $\tau_t \simeq 1.2$. 
The $\tau_t=1.2(1)$ is out of the range of neuro experiments:
$1.5 < \tau_t < 2.4$ \cite{brainexp}, but a good agreement/overlap can be found
in the sub-threshold region. To investigate the effects of inhibition as
in \cite{CCdyncikk} here we also studied the modified {\it KKI-18} graph,
in which we flipped a small fraction of weight links randomly.

\subsubsection*{Inhibitory weights}

\begin{figure}[!h]
  \includegraphics[height=5.5cm]{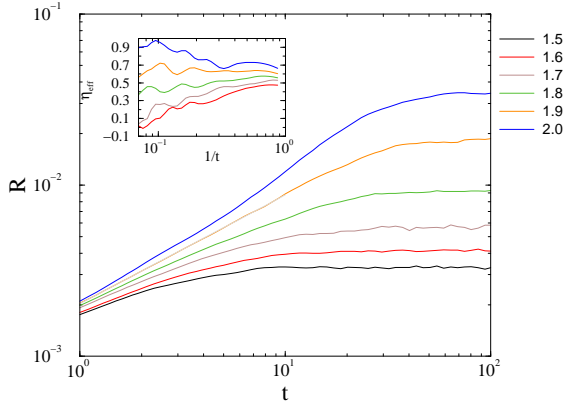}
\caption{Growth of the average $R$ on the inhibitory {\it KKI-18-I} graph 
near the synchronization transition point for: $K=1.5$, $1.6$, $1.7$, 
$1.8$, $1.9$, $2.0$ (bottom to top curves). 
Inset: effective exponents, defined by (\ref{Reff}) for the same curves.}
\label{figccIgrowth}
\end{figure}

\begin{figure}[!h]
 \includegraphics[height=5.5cm]{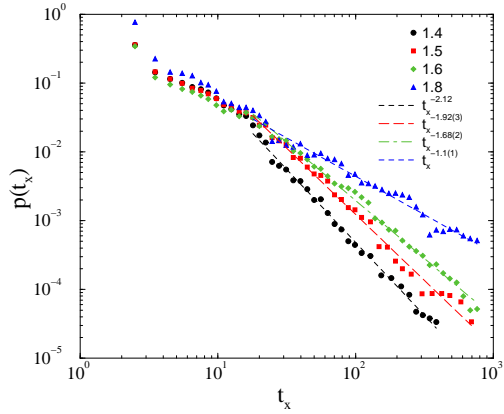}
\caption{Duration distribution of $t_x$ on the {\it KKI-18-I} model for growth
$K=1.4$ (bullets), $1.5$ (boxes), $1.6$ (diamonds), $1.8$ (triangles).
The dashed line shows PL fits to the tail region: $t_x > 20$.}
\label{durcci}
\end{figure}

We repeated the analysis of the previous section for the {\it KKI-18-I} graph, 
possessing $5\%$ inhibitory link fraction symmetrically : 
$W''_{ij} = W''_{ji} = -W'_{ij}$.
First we located the transition point as shown on Fig.~\ref{figccIgrowth}.
The crossover to synchronization occurs at $K_c = 1.9(1)$, slightly higher 
than in case of the {\it KKI-18} network.
The tails of the $p(t_x)$ probability distributions exhibit PL-s with 
$1 < \tau_t \le 2$ in the $1.4 < K < 1.8$ region.
These exponent values overlap the range of experiments (see Fig.~\ref{durcci}).
The large variation of $\tau_t$ below the transition, suggests the
strong disorder may cause some GP effects, but in the lack of true 
phases we can only claim resemblance with recent results on 
power-grid networks~\cite{POWcikk},
where we pointed out a relation to a phenomena called 
frustrated synchronization~\cite{Frus,FrusB}. 

We have redone this analysis for another random $5\%$ inhibitory link
graph sample and arrived to the same results. In \cite{CCdyncikk} the
robustness of GP in case of the threshold model dynamical behavior has
been tested by the random neglection of 20\% of links in one direction.
Here we considered the situation of the neglection of all links in one
direction: $W''_{ij} = -W'_{ij}$,  $W''_{ji} = 0$. Even in this 
extreme anisotropic case one case find a similar extended scaling 
region below a very smooth crossover as shown in the Supplementary
material. 

\begin{figure}[!h]
\includegraphics[height=5.5cm]{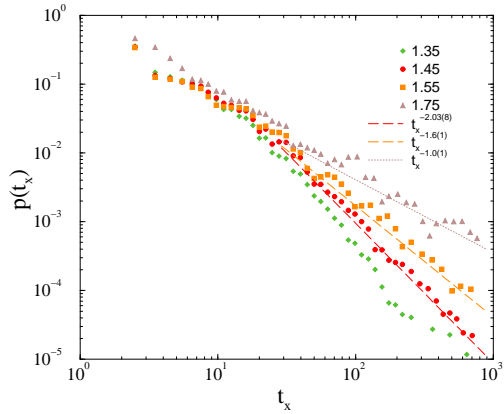}
\caption{Duration distribution of $t_x$ on the {\it KKI-18-I} model 
in case of $5\%$ inhibitory node assumption for $K=1.35$ (+),
$K=1.45$ (bullets), $1.55$ (boxes), $1.75$ (triangles).
The dashed line shows PL fits to the tail region: $t_x > 20$.}
\label{figInode5}
\end{figure}

Finally, we considered graphs with $5, 10, 20\%$ inhibitory node
assumptions, by flipping signs of (out or in) link weights of these
randomly selected sites. Note, that nodes here represent big bunches 
of neurons. We show the duration distribution results here for one of 
the $5\%$ inhibitory node case, when signs of out links are reversed.
Below the synchronization transition point, which is at $K_c=1.7(1)$
we can find again a region : $1.35 < K < 1.7$, where PL tailed
de-synchronization duration durations emerge as before 
(see Fig.~\ref{figInode5}), characterized by the exponents $1 < \tau_t < 2$.

We have also studied the synchronization behavior on graphs with 
fully inhibited node cases of $5, 10, 20\%$, by flipping the weight 
signs of in-links. Even without weight normalization this produces
crossovers at $K_c  \simeq 0.18$ with $\eta_\mathrm{eff} \simeq 0.23(5)$,
different from the $\eta \simeq 0.6$ value we found up to now.
The tails of the $p(t_x)$ probability distributions exhibit PL-s with
well inside the range of experiments (see Supplementary material).

\section*{Conclusion and discussion}

Brain experiments support evidence for power-law distributed activity avalanches.
These have been explained mainly by discrete, threshold type of models,
showing exponents close to the MF values, in agreement with neuro measurements.
Oscillatory activity is widespread in dynamic neuronal networks \cite{PSM16}.
The Blue brain project \cite{Ma06} suggests that the cortical dynamics 
operates at the edge of a phase transition between an asynchronous phase 
and a synchronous one with emerging oscillations \cite{Ma15}.
A recent MF theory showed the emergence of scale-free avalanches 
at the edge of synchronization \cite{MunPNAS}. 

An extension, taking into account network heterogenities of a large human 
connectome is provided here, within the framework of the Kuramoto model.
We determined the phase synchronization transition points and provided 
characteristic time exponents, which describe synchronization or 
de-synchronization events. 
We found good agreement with the neuro-experimental values in an extended
range below the crossover point. We investigated the case, when 
$5, 10, 20\%$ fraction of the randomly selected links or nodes are 
set inhibitory. 
The obtained $\tau_t$ duration exponents have been found to
be invariant for these fractions and are in the range of experiments.
This conclusion has already been derived within the framework of discrete 
threshold models \cite{CCdyncikk,CC-tdepcikk}.

At the transition point the Kuramoto model on these homeostatic 
{\it KKI-18} network exhibits: $\tau_t = 1.2(1)$, well below the 
results for the 2dll graph, $\tau_t = 1.6(1)$, which is expected 
to be a system of MF interactions. 
However, even for this MF like model the dynamical exponents were found to 
be slightly away from the Kuramoto MF values \cite{cmk2016}. 
This can be the result of enormous corrections to scaling, or due to
quenched heterogeneity effects of the 2dll graph. The details of this 
problem is discussed in \cite{Kurcikk}. The effective growth exponent 
on the investigated connectome networks is $\eta \simeq 0.6$, near to 
the 2dll graph results, except for the node inhibited case, where
$\eta_\mathrm{eff} \simeq 0.25(5)$ has been found.

Although in the {\it KKI-18} graph the topological dimension is below
$d_l=4$, a crossover behavior can clearly be identified. Around
this smeared transition we found scale-free de-synchronization 
'avalanche' tails, like in case of the dynamical criticality of
GP, pointing out relation to possible frustrated synchronization 
effects~\cite{Frus,FrusB}. Modules of the connectome graph enhance 
rare-region effects or frustrated synchroniztion domains.

As it was discussed in \cite{CCdyncikk} such coarse grained connectomes 
suffer possible sources of errors, like unknown noise in the data generation;
underestimation of long connections; radial accuracy, influencing
endpoints of the tracts and hierarchical levels of the cortical organization;
or transverse accuracy, determining which cortical area is connected
to another. Still important modifications, such as inhibitory links,
directedness, or random loss of connections up to $20\%$ confirmed
the robustness of dynamical scaling, suggesting that fine network details
may not play an important role. It is also important, that the PL tail in
the weight distribution is similar to what was obtained by a synaptic
learning algorithm in an artificial neural network \cite{scarp18}.
A very recent experimental study has provided confirmation for the 
connectome generation used here \cite{newcomp}. 
Those results suggest that diffusion MRI tractography is a powerful tool 
for exploring the structural connectional architecture of the brain.  

>From a neuroscience point of view one may find the Kuramoto model too 
simplistic to describe the brain. However, at least in the weak
coupling limit equivalence of phase-oscillator and integrate-and-fire 
models has been found \cite{PolRos15}, that may hold for the
sub-critical region, where we observed the dynamical scaling.
At criticality universal scaling is expected to hold and thus 
the exponents obtained for this model can well describe the 
synchronization transition of other, more complex models. 
We believe that investigating a basic model
is an important and necessary first step for neuroscience as 
this can provide a representative for a whole class of more real ones.
 
An interesting continuation of this work would be the study of
the frequency entrainment of oscillators, which can exhibit
real phase transition at $d=3.05$ of the KKI-18 graph, or 
consideration of more complex models or graphs than the ones 
investigated here.
Another open point could be the determination of avalanche sizes
in these synchronization processes.
The codes and the graphs used here are available on request from the 
corresponding author.

\section*{Acknowledgments}
We thank R. Juh\'asz for the useful comments and Wesley Cota for
generating and providing us Fig. 1. 
We gratefully acknowledge computational resources provided by NIIF Hungary, the
HZDR computing center, the Group of M. Bussmann and the Center for Information
Services and High Performance Computing (ZIH) at TU Dresden via the GPU Center
of Excellence Dresden. We thank S. Gemming for support.
Support from the Hungarian research fund OTKA (K128989), the
Initiative and Networking Fund of the Helmholtz Association via the W2/W3
Programme \mbox{(W2/W3-026)} and the Helmholtz Excellence Network
DCM-MatDNA (ExNet-0028) is acknowledged.

\bibliography{bib}

\end{document}